\documentclass{emulateapj}
\usepackage{natbib}

\newcommand{\msun}{\ensuremath{{\rm M}_\odot}}

\begin{document}

\journalinfo{Accepted by the Astrophysical Journal}

\title{Masses, Parallax, and Relativistic Timing of the 
PSR~J1713+0747 Binary System}

\author{Eric M. Splaver and David J. Nice}
\affil{Physics Department, Princeton University \\ 
Box 708, Princeton, NJ  08544}

\author{Ingrid H. Stairs}
\affil{Department of Physics and Astronomy, University of British Columbia\\
6224 Agricultural Road, Vancouver, BC V6T 1Z1, Canada}

\author{Andrea N. Lommen}
\affil{Department of Physics and Astronomy, Franklin and Marshall College \\
Box 3003, Lancaster, PA 17604}

\and

\author{Donald C. Backer}
\affil{Department of Astronomy and Radio Astronomy Laboratory, 
University of California, Berkeley, CA 94720}

\medskip

\submitted{Accepted for publication by the Astrophysical Journal}

\begin{abstract}

We report on 12 years of observations of PSR~J1713+0747, a pulsar in a 68-day
orbit with a white dwarf.  Pulse times of arrival
were measured with uncertainties as small as 200\,ns.  
The timing data yielded measurements of
the relativistic Shapiro delay, perturbations of pulsar orbital
elements due to secular and annual motion of the Earth, and the
pulsar's parallax, as well as pulse spin-down, astrometric,
and Keplerian measurements. 
The observations constrain the masses
of the pulsar and secondary star to be $m_1=1.3\pm 0.2\,\msun$
and $m_2=0.28\pm 0.03\,\msun$, respectively (68\% confidence).
Combining the theoretical orbital period-core mass relation with
the observational constraints yields a somewhat higher pulsar mass,
$m_1=1.53_{-0.06}^{+0.08}\,\msun$.  The parallax is $\pi=0.89\pm 0.08$\,mas,
corresponding to a distance of $1.1\pm 0.1$\,kpc;  the precision of
the parallax measurement is limited by uncertainties in the electron
content of the solar wind.  The transverse velocity is unusually small, 
$33\pm3$\,km\,s$^{-1}$.  We find significant timing noise on time
scales of several years, but no more than expected by extrapolating
timing noise statistics from the slow pulsar population.  With the
orientation of the binary orbit fully measured, we are able to
improve on previous tests of equivalence principle violations.

\end{abstract}


\section{Introduction}\label{sec:intro}

High precision timing of a radio pulsar binary
system can reveal a wealth of information about the dynamics
of the binary, the astrometry of the system, and the natures
of the pulsar and secondary stars.  
PSR~J1713+0747, a 4.6\,ms pulsar in a 67.8~day orbit with
a white dwarf, is among 
the very best pulsars for high precision timing:  it has a 
high flux density, shallow spectrum, and sharp pulse peak
\citep{fwc93}.  

We observed PSR~J1713+0747 over six years as part of
a systematic high precision pulsar timing study at the
Arecibo Observatory.  We used data acquisition
systems employing coherent dedispersion, allowing substantial
improvements in timing precision compared to previous work.

Early observations of this source were reported by
\cite{cfw94} (herein CFW), who analyzed 1.5 years of pulsar
timing data.  Previous observations have also been reported
by \cite{vb03}.  
We combined the data of CFW with our newer observations
to produce a single data set
spanning twelve years.  The superior timing precision of the
new data and the longer total time span of observations
yield substantial refinements of all measurements reported by CFW, as
well as a number of additional new measurements.

High precision detection of the Shapiro delay allows the pulsar
and secondary star masses to be separately measured.  
The Shapiro delay measurement reported in CFW did not have sufficient
precision to independently determine the pulsar and secondary star masses.
These quantities are of interest because, while measured double
neutron star
masses fall into a very narrow range, 1.25 to 1.44\,\msun\ 
\citep{lp04}, extended accretion during the formation of pulsar--white dwarf
binaries may lead to higher pulsar masses in these systems.

High precision timing of pulsars can be used to place limits
on the gravitational wave background at frequencies of
$10^{-6}$ to $10^{-5}$ Hz by searching for arrival time variations
on time scales of years \citep{lbsn03,ktr94},
but such measurements require that the timing signal not be
contaminated by ``timing noise,'' random variations of the
pulsar rotation.  Timing noise is inversely correlated with
rotation period derivative \citep{antt94}, so millisecond pulsars,
which have very small period derivatives, are prime candidates
for gravitational wave studies.  The 12-year time span of
data on PSR~J1713+0747 has revealed
significant pulse arrival variations beyond those expected
from simple magnetic dipole spin-down.  We quantify
this apparent timing noise, but we leave the analysis of
the PSR~J1713+0747 signal in the context of the
gravitational wave background to another work.

We detected both annual and secular perturbations of
orbital elements due to the changing Earth--binary line-of-sight.
This allowed us to uniquely determine the inclination and position
angle of the orbit.  This is
only the second pulsar binary system for which timing observations
have fully specified both angles of the orbital orientation
\citep{vbb+01,vb03}.

Preliminary results from this project, using data collected
through mid-2002, were published elsewhere \citep[e.g.,][]{nss03,nss04}.  
The present paper incorporates two further years of high quality
data, resulting in some changes to the best-fit parameters, but 
within the expected uncertainties.

The plan of this paper is as follows.  In \S 2 we summarize the observations.
In \S 3 we describe the timing model.  In \S 4 we present the parameters of
the pulsar and binary system derived from fitting the timing model to the 
observational data.  In \S 5 we analyze the timing model parameters in
the contexts of timing noise, stellar masses, pulsar distance and velocity,
and theories of strong field gravitation.  In \S 6 we summarize the
key results.

\section{Observations}\label{sec:obs}

\begin{deluxetable*}{ccccccc}[t]
\tablewidth{0pt}
\tablecaption{Summary of Observations\label{tab:obs}}
\tablehead{
System       & Dates           &  Frequency & Bandwidth        & Number & Typical             & RMS           \\
             &                 &    (MHz)   &  (MHz)           & of     & Integration         & Residual$^a$  \\
             &                 &            &                  & TOAs   & (min)               & ($\mu$s)      \\
}
\startdata
Mark III     &  1992.6--1993.0 &  1400      &  \phn40          & \phn\phn9  &   47                &  \phn 0.70$^b$\\
             &  1992.3--1994.1 &  1400      &  \phn40          & \phn59     &   47                &  \phn 0.41$^c$\\
ABPP         &  1998.1--2004.4 &  1410      &  \phn56          & 101     &   60                &  0.18         \\
             &  1999.7--2004.4 &  2380      &  112             & \phn49     &   30                &  0.35         \\
Mark IV      &  1998.6--2004.4 &  1410      &  \phn10          & \phn81     &   58                &  0.28         \\
             &  1999.8--2004.4 &  2380      &  \phn10          & \phn44     &   29                &  0.46         \\
\enddata
\tablenotetext{a}{Values incorporate the effect of averaging TOAs from shorter integration times.}
\tablenotetext{b}{Filter bank used a 78-$\mu$s time constant.}
\tablenotetext{c}{Filter bank used a 20-$\mu$s time constant.}
\end{deluxetable*}

The 305-m Arecibo radio telescope recorded a total of 343 pulse times of
arrival (TOAs) of PSR~J1713+0747 on 166 separate days between April 1992
and May 2004.  Table~\ref{tab:obs} summarizes the observations.  A shutdown
of the telescope for an upgrade resulted in a gap in the 
observations between 1994 and 1998.

Observations made in 1992 and 1993 are described in CFW.  Briefly, those
data consist of 68 TOAs collected at intervals of about two weeks using the
Princeton Mark~III observing system \citep{skn+92} coupled to a $2 \times
32 \times 1.25$\,MHz filter bank, with 32 spectral channels for each sense
of polarization.  In every scan, the square-law detected outputs of each
pair of channels (representing two polarizations) were smoothed with 
time constants of 20\,$\mu$s or 78\,$\mu$s, summed, folded synchronously
at the predicted pulse period, and shifted in time to remove dispersive
delay. The channels were then summed to create total-intensity pulse
profiles of 256 bins.

From 1998 through 2004, 
observations were made using
Princeton Mark~IV system \citep{sst+00} and the Arecibo-Berkeley Pulsar
Processor (ABPP), often running in parallel to
analyze the same radio frequency signal.
A total of 275 TOAs were obtained on 132 different
days, usually at intervals of a few weeks, but occasionally more
densely sampled.  A typical day included an hour of
observations at 1410\,MHz and a half hour of observations at 2380\,MHz.  

The Mark~IV system critically samples and records 10 MHz pass bands in each
sense of circular polarization, quantized with 2-bit resolution.  The recorded
voltages are analyzed off line: the data stream is coherently dedispersed,
after which the self- and cross-products of the voltages are calculated and folded synchronously at the pulse
period.  Observations are continuous over (typically) 29 minute intervals, but
are analyzed in blocks of 190 seconds, each of which yields a 1024 bin pulse
profile with four Stokes parameters.

The ABPP filters the passband into narrow spectral channels,
samples voltages with 2-bit resolution, 
and applies coherent dedispersion to each channel using
3-bit coefficients.
For PSR~J1713+0747 at 1410\,MHz, thirty-two spectral channels of
width 1.75\,MHz are processed in each polarization,
for a total bandwidth of 56\,MHz;
at 2380\,MHz, the channel bandwidths are 3.5\,MHz, for
a total bandwidth of 112\,MHz.
The dedispersed time series in each channel
is folded synchronously at the pulse period and integrated
for 180 seconds.

We used conventional techniques to measure pulse arrival times.  For
the Mark~III and Mark~IV data, each pulse profile was cross-correlated
with a standard template to measure the phase offset of the pulse
within the profile.  For the ABPP data, an analytical model of the
pulse profile was fit to each data profile.  In either case, the time
offset measured from the profile was added to the start time of the
integration and translated to its middle to yield a TOA. The start
times were referenced to the observatory atomic clock, which was
corrected retroactively to the UTC timescale using data from the
Global Positioning System (GPS) satellites. For each data acquisition
system, all TOAs at a given frequency collected on a given day were
averaged to make a single effective TOA.  The time intervals
spanned by these average points is listed as ``typical integration''
in Table~\ref{tab:obs}.

The ABPP and Mark~IV systems often ran in parallel, analyzing the same radio
frequency signal processed through the same amplifiers and many
of the same filters.  Of the TOAs
summarized in Table~\ref{tab:obs}, seventy-six
 pairs of ABPP and Mark~IV TOAs were
collected simultaneously.  Nevertheless, we have treated them as independent data
streams in the timing analysis.  There are two justifications for this
approach.  First, the width of the passband measured by the ABPP is
substantially larger than that measured by Mark~IV, so the signal measured by
the two machines are somewhat different.  Second, empirical tests of the
residual arrival times, after removing the best fit timing model, show
only modest correlation between TOAs measured by the two systems.
Correlation coefficients are 0.33 and 0.30 for the 1410 and
2380\,MHz residual TOAs, respectively.  

We found that 
formal measurement uncertainties calculated directly from the
TOA measurements of individual data records tended to
moderately underestimate the true scatter in the arrival times.  The
cause of the underestimation is not known, but it is
a common phenomenon in millisecond pulsar timing.  
A particular challenge for these observations is 
the correction of the coarse quantization of incoming signals given
the highly variable nature of the signals.  
In any case, we added
``systematic'' terms in quadrature to these formal uncertainties
in order to produce timing fits with reduced $\chi^2$ values close to 1.  
The root-mean-square
(RMS) residual arrival times listed in Table~\ref{tab:obs} were
calculated by giving all TOAs equal weight, independent of uncertainties.

\begin{figure}[b]
\plotone{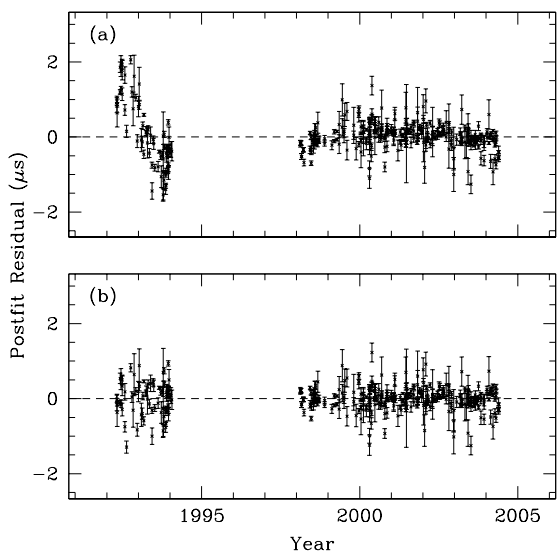}
\caption{Post-fit residuals of
PSR~J1713+0747. (a) Pulsar rotation modeled by $\nu_0$ 
and $\nu_1$ only; timing noise is evident.  Because the
early data (pre-1995) are less precise than the later
data, they are down-weighted in the fit;  as a result,
the fit allows
their residual arrival times show greater variation.
Alternate values of $\nu_0$ and $\nu_1$ could
be found which reduced the timing noise in the early
data at the expense of increasing it in the later data. 
(b) Pulsar rotation modeled by
$\nu_0$ through $\nu_8$.  Error bars shown here do not include compensation
for systematic uncertainties, although such compensation was included
in these timing models.
\label{fig:res}}
\end{figure}

\section{Timing Model}\label{sec:narrative}

We used the {\sc tempo}\footnote{http://pulsar.princeton.edu/tempo} software
package to fit a pulse timing model to the observed TOAs.  The model
incorporated pulsar rotation, astrometry, orbital motion, and dispersion of
the pulsar signal by the ionized interstellar medium.  The fits allowed for
arbitrary time offsets between sets of TOAs taken with different observing
systems and at different frequencies.  Earth motion was modeled using the JPL
DE405 ephemeris \citep[][see also \S\ref{sec:eph}]{sta98,sta04b}. 
The ultimate time reference was 
the TT(BIPM03) scale \citep{bipm03} adjusted to barycenter time scale
TDB with the TE405 time ephemeris \citep{if99}.

In this section, we describe the elements of the timing model in detail.
We defer discussion of the results of the model fit until
\S\ref{sec:grid}.  However,
we will make reference of these results to justify
components of the timing model.

\subsection{Pulsar Rotation and Timing Noise}\label{sec:rotation}

The pulsar rotation frequency at time $t$ can be written as a polynomial
expansion,
$\nu(t) = \nu_0 + \nu_1(t-t_0)+\frac{1}{2}\nu_2(t-t_0)^2+...$, where $t_0$ is
a reference epoch near the center of the data span and $\nu_0$, $\nu_1$,
$\nu_2$, etc., are the pulsar rotation frequency, frequency derivative,
frequency second derivative, and so on.  Under the standard model of a neutron
star with a rotating magnetic dipole, a millisecond pulsar such as
PSR~J1713+0747
should have negligibly small values of $\nu_2$ and higher order
terms.   However, timing
fits incorporating only $\nu_0$ and $\nu_1$ leave a
systematic signature in the residual arrival times after
removing the best-fit model (Figure~\ref{fig:res}).  To
remove the long-term systematic trend, we ``whitened'' the
data by incorporating seven additional terms to the model, $\nu_2$ through $\nu_8$.
These terms correspond
to time scales on the order of years, much longer than the orbital
period of the pulsar, and hence they have negligible effect on the 
measurement of orbital elements.

\subsection{Dispersion Measure Variations}\label{sec:dm}

Radio pulses traversing the solar system and the interstellar
medium are delayed by dispersion. The time delay in seconds is 
$\Delta t_{\rm DM}={\rm{DM}}\,/ (2.41 \times 10^{-16}\,f^2)$, where $f$ is the
radio frequency in Hz
and the integral of electron density along the line of sight,
${\rm DM}=\int n_e(l)\,dl$, is the
dispersion measure in pc\,cm$^{-3}$.  Observations at
two frequencies, 1410 and 2380 MHz, over the last 5 years of observations,
allow us to search for variations in DM over time.  
Of particular concern are variations due to propagation through
the ionized solar wind as the Earth moves about the Sun.
We modeled the solar electron density at distance
$r$ from the Sun as $n_e(r)=n_0(1\,\mbox{\sc au}/r)^2$, where $n_0$ is the 
electron density at r=1\,{\sc AU} which we take as a free parameter
in the pulsar timing solution.  The solar contribution to ${\rm DM}$ 
is calculated by integrating $n_e(r)$ along the path from Earth toward
the pulsar.  Ulysses data have
shown the $1/r^2$ scaling to hold 
over a wide range of heliocentric latitudes \citep{ihmm01}.  
The $1/r^2$ model is clearly an oversimplification---it neglects the substantial
difference between the high density slow wind along the ecliptic
and the low density wind at high latitudes, and it does not
allow for temporal variations in $n_0$.  However, because it has
a simple analytic form and requires only a single additional 
parameter in the fit, it is a convenient form to use for the timing
model.
The best fit electron density at 1\,{\sc au} is 
$n_0=5\pm 4$\,electrons per cm$^3$.  This gives peak-to-peak
DM variations of 0.0002\,pc\,cm$^{-3}$, and arrival time variations
of 400\,ns at 1410\,MHz.

\begin{figure}[b]
\plotone{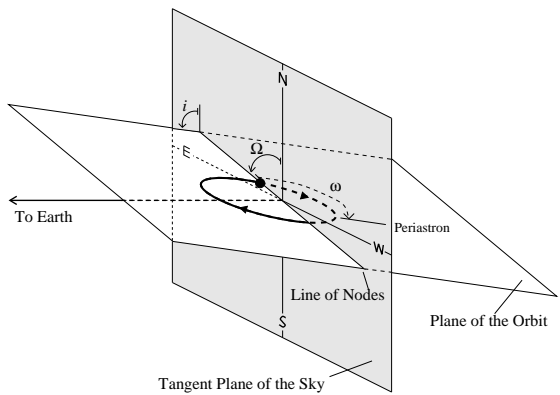}
\caption{
Geometry of the binary system, showing position angle 
of ascending node, $\Omega$, inclination, $i$, and angle of
periastron, $\omega$.  The black dot indicates the ascending
node.
\label{fig:angles}}
\end{figure}

\begin{figure}[b]
\plotone{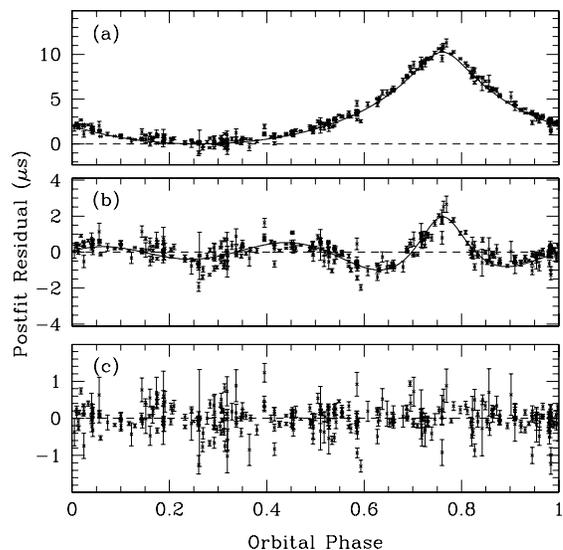}
\caption{Shapiro delay in the timing residuals of PSR~J1713+0747
as a function of orbital phase.  (a)  The Shapiro delay
term is excluded from the model while the rest of the parameters
are fixed at their best-fit values from
Table~\ref{tab:param}.
The solid curve shows the general relativistic
delay predicted by equation~\ref{eqn:shapiro}. 
(b) The Shapiro
term is excluded from the model but the other
parameters are permitted to vary. 
(c)  Shapiro delay is included in the model.
The residuals in this panel are from the same
timing fit as those in Figure~\ref{fig:res}b.} 
\label{fig:shapiro}
\end{figure}

\subsection{Orbital Kinematics Including Shapiro Delay}\label{sec:shapiro}

Orbital kinematics were incorporated into the timing model
by means of the theory-independent representation of
Damour \& Deruelle (1986)\nocite{dd86}.  
Parameters of the orbital model include (1)
five Keplerian orbital elements:
orbital period, $P_b$;
semi-major axis projected into the line of sight,
$x=(a_1\sin i)/c$, where $a_1$ is the semi-major axis,
$i$ the inclination angle, and $c$ the speed of light;
eccentricity, $e$; angle of periastron, $\omega$;
and time of periastron passage, $T_0$;  (2) secular 
variations of the Keplerian elements, most notably
the time derivative of $x$, denoted $\dot{x}$; 
(3) the orientation of the system, defined by the
inclination of the orbit, $i$, where $0^\circ \le i < 180^\circ$,
and the position angle of ascending node, $\Omega$, 
where $0^\circ \le \Omega < 360^\circ$ and
$\Omega$ is defined north through east\footnote{For
celestial position angles, we follow the standard convention that
$0^\circ$ is north and $90^\circ$ is east.  For inclination,
we follow the convention that an orbit with $i=0^\circ$ has an
angular momentum vector pointing toward Earth. 
These definitions differs 
from Kopeikin (1995, 1996)\nocite{kop95,kop96} 
and \cite{vbb+01}, who define $\Omega$ such that
$0^\circ$ is east and $90^\circ$ is north, and define
$i=0^\circ$ to be an orbit with angular momentum vector
pointing away from Earth.
\label{foot:omega}}
(see Figure~\ref{fig:angles}); and (4) the masses
of the pulsar and the secondary star,
$m_1$ and $m_2$.

The masses and inclination are connected by the mass function,
\begin{equation}
f_1\equiv\frac{(m_2 \sin i)^3}{(m_1+m_2)^2}
        = \frac{x^3}{T_\odot}\left(\frac{2\pi}{P_b}\right)^2,
\label{eqn:massfunc}
\end{equation}
where $T_{\odot}\!=\!GM_\odot/c^3\!=\!4.925\!\times\!10^{-6}\,$s.
In the analysis below, we treat $i$ and $m_2$ as independent parameters
in the timing model, and we use equation~\ref{eqn:massfunc} to determine $m_1$.

According to general relativity, pulses are retarded as they propagate through
the gravitational potential well of the secondary.  For a nearly circular
orbit, this Shapiro delay is
\begin{equation}
\Delta t_s = -2\,m_2\,T_\odot \ln[1-\sin{i} \sin(\phi+\omega)],
\label{eqn:shapiro}
\end{equation}
where $\phi$ is the orbital phase.  In principle,
measurement of the Shapiro delay yields $m_2$ and $\sin i$.
In practice, unless $\sin i\sim 1$, the Shapiro delay is highly
covariant with $x$ in the timing fit and hence difficult to measure.
Nevertheless, as shown in Figure~\ref{fig:shapiro}, the Shapiro delay
can be clearly distinguished in the PSR~J1713+0747 data.  
The constraints
on $m_2$ and $i$ that arise from this measurement are quantified in
\S\ref{sec:grid}.  Since $0^\circ\le i<180^\circ$ (see
Figure~\ref{fig:angles}), there
in an ambiguity in deriving $i$ from $\sin i$ in equation~\ref{eqn:shapiro},
because both $i$ and $180^\circ\!-\!i$ are solutions.  The resolution
of this ambiguity is discussed below.

\subsection{Projection Effects due to Proper Motion}\label{sec:xdot}

Proper motion of a binary system results in secular changes
in $\omega$ and $x$ \citep{kop96}.  For a nearly circular orbit,
a secular change in $\omega$ is indistinguishable from a small
perturbation of the orbital period, and hence is unmeasurable.
In contrast, the secular change in $x$ 
is significant.  The time derivative of $x$, $\dot{x}$, is given 
by
\begin{equation}
\frac{\dot{x}}{x}=\mu\,\cot i\,\sin(\theta_\mu-\Omega),
\label{eqn:xdot}
\end{equation}
where $\mu$ and $\theta_\mu$ are the magnitude and position angle 
of the proper motion, respectively.
In effect, $i$ is determined by the Shapiro delay
(\S\ref{sec:shapiro}), and $\Omega$ is then constrained by the $\dot{x}$
measurement.  
Solving equation~\ref{eqn:xdot} for $\Omega$ yields two possible
values for each of the two values of $i$
(\S\ref{sec:shapiro}).
Thus there are four
distinct binary orientations (combinations of $i$ and $\Omega$)  
allowed by the Shapiro delay and $\dot{x}$
measurements.

\subsection{Annual-Orbital Parallax}\label{sec:annorbpx}

The Earth's annual motion about the solar system barycenter
changes the line-of-sight to the binary system and perturbs
the binary parameters $x$ and $\omega$, an effect known
as annual-orbital parallax \citep{kop95,vbb+01,vb03}.  
We incorporated annual-orbital parallax
into the timing model by perturbing
the values of $x$ and $\omega$ before calculating pulse arrival
time delays using a standard orbital model.  The perturbation
formulae are given 
by Kopeikin (1995)\nocite{kop95}; we summarize them here.
The position of the Earth relative to the solar system
barycenter is given by the vector X, Y, Z. The pulsar
position is right ascension $\alpha$ and declination $\delta$. 
Define
\begin{equation}
\begin{array}{c}
\Delta_{\rm I0} =-X\sin\alpha+Y\cos\alpha, \\
\Delta_{\rm J0} =-X\sin\delta\cos\alpha-Y\sin\delta\sin\alpha
             +Z\cos\delta.
\end{array}
\end{equation}
These are the east and north motions of the Earth in a coordinate
system parallel to the plane of the sky.
The observed values of $x$ and $\omega$ are perturbed from
their intrinsic values 
by
\begin{equation}
x_{\rm obs} = x_{\rm int}\left[1-\frac{\cot i}{d}(\Delta_{\rm I0}\cos\Omega-\Delta_{\rm J0}\sin\Omega)\right], 
\label{eqn:annorbx}
\end{equation}
and
\begin{equation}
\omega_{\rm obs} = \omega_{\rm int} - \frac{\csc i}{d}(\Delta_{\rm I0}\sin\Omega+\Delta_{\rm J0}\cos\Omega).
\label{eqn:annorbomega}
\end{equation}

\begin{figure}[b]
\plotone{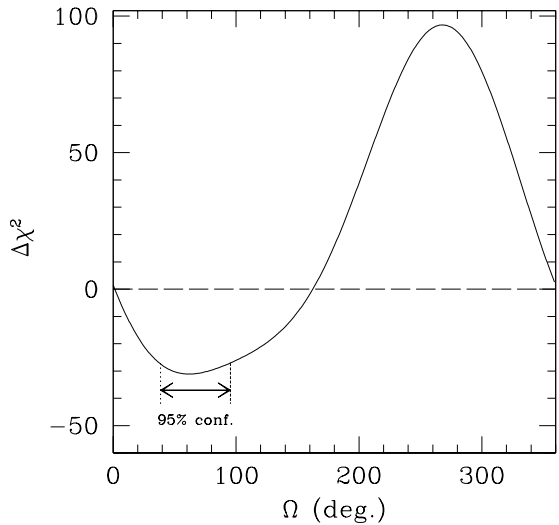}
\caption{The effect of annual-orbital parallax on the timing
solution is shown by displaying the
difference in goodness of fit, $\Delta\chi^2$, as a function
of position angle of ascending node.  The value $\Delta\chi^2=0$ 
corresponds to the best timing model without annual-orbital
perturbations.  The 95\% confidence range of $\Omega$
is indicated.
\label{fig:annorb}
}
\end{figure}

We found the annual-orbital perturbation of $x$ to have only
a marginal effect on the timing of PSR J1713+0747, and no
useful measurements can be derived from it.  (Nevertheless,
the $x$ perturbation was included in the timing model, since
it introduced no additional degrees of freedom.)  On the other
hand, the perturbations of $\omega$ are sufficiently large that
incorporating them into the timing model significantly improves
the goodness of the timing fit. Figure~\ref{fig:annorb} shows
the $\chi^2$ values for fits with and without the annual-orbital
parallax perturbations.  In these fits, $m_1$, $i$, $\dot{x}$, 
and $\Omega$ were treated as independent parameters, unconstrained by
equation~\ref{eqn:xdot}; hence the best-fit range of $\Omega$
in Figure~\ref{fig:annorb} differs somewhat from that given in \S 4.

As a practical matter, for PSR~J1713+0747 the annual-orbital parallax 
does little to improve the precision of
the timing parameter measurements (since $i$ and $\Omega$ are
better measured by Shapiro delay and $\dot{x}$). However, it 
allows the fourfold ambiguity in $i$ and $\Omega$ to be broken,
picking out one distinct orientation of the orbit.

\begin{figure}[b]
\plotone{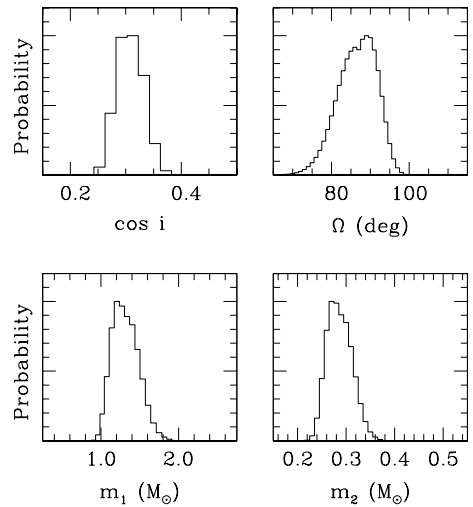}
\caption{Probability distribution functions for values
of $\Omega$, $m_1$, $m_2$, and $\cos i$.}
\label{fig:pdfs}
\end{figure}

\begin{deluxetable*}{ll}[t]
\tablewidth{0pt}
\tablecaption{Timing Model Parameters\tablenotemark{a}\label{tab:param}}
\tablehead{\multicolumn{2}{c}{Measured Quantities}}

\startdata
Right Ascension, $\alpha$ (J2000)\dotfill                           & $17^{\rm h}13^{\rm m}49\fs5305335(6)$\\
Declination, $\delta$ (J2000)\dotfill                               & $+07^\circ47\arcmin37\farcs52636(2)$\\
Total proper motion, $\mu$ (mas yr$^{-1}$)\dotfill                  & 6.297(7) \\
Position angle of proper motion, $\theta_\mu$\dotfill               & 128\fdg66(7) \\
Parallax, $\pi$ (mas)\dotfill                                       & 0.89(8)\\
Rotation frequency, $\nu_0$ (s$^{-1}$)\dotfill                      & 218.8118439157321(3)\\
First derivative of $\nu_0$, $\nu_1$ (s$^{-2}$)\dotfill               & $-4.0835(2)\times 10^{-16}$ \\
Epoch, $t_0$ (MJD [TDB])\dotfill                                    & 52000.0 \\
Dispersion Measure, DM$_0$ (pc\,cm$^{-3}$)\dotfill                  & 15.9960 \\
Orbital period, $P_b$ (days)\dotfill                                & 67.8251298718(5)\tablenotemark{b}\\
Projected semi-major axis, $x$ (lt-s)\dotfill                       & 32.34242099(2)\tablenotemark{b}\\
Eccentricity, $e$\dotfill                                           & 0.0000749406(13)\tablenotemark{b}\\
Time of periastron passage, $T_0$ (MJD [TDB])\dots                  & 51997.5784(2)\tablenotemark{b}\\
Angle of periastron, $\omega$ (deg)\dotfill                         & 176.1915(10)\tablenotemark{b}\\
Cosine of inclination angle, $\cos i$\dotfill                       & $0.31(3)$\\
Position angle of ascending node, $\Omega$ (deg)\dotfill            & $87(6)$\\
Companion mass, $m_2$ (M$_\odot$)\dotfill                           & $0.28(3)$ \\

\cutinhead{Measured Upper Limits}

First derivative of DM, DM$_1$ (pc\,cm$^{-3}$\,yr$^{-1}$)\dotfill        & $0(2)\times 10^{-5}$ \\
First derivative of $\omega$,  $\dot{\omega}$ (deg\,yr$^{-1}$)\dotfill   & $6(4)\times 10^{-4}$ \\
First derivative of ${P_b}$, $\dot{P}_b$ \dotfill                        & $0(6)\times 10^{-13}$ \\

\cutinhead{Derived Quantities}

Proper motion in $\alpha$, $\mu_{\alpha}=\dot{\alpha}\cos\delta$ (mas yr$^{-1}$)\dotfill
                                  & 4.917(4)\\
Proper motion in $\delta$, $\mu_\delta=\dot{\delta}$ (mas yr$^{-1}$)\dotfill
                                  & $-$3.933(10)\\
Galactic longitude, $l$\dotfill                       & $28\fdg 748$\\
Galactic latitude, $b$\dotfill                        & $25\fdg 222$\\
Distance, $d$ (kpc)\dotfill                           & $1.1(1)$\\
Rotation period, ${P}$ (s)\dotfill                 & 0.004570136525082781(6)\\
Observed rotation period derivative, $\dot{P}_{\rm obs}$\dotfill       & $8.5288(3)\times 10^{-21}$\\
Intrinsic rotation period derivative, $\dot{P}_{\rm int}$\dotfill       & $8.1\times 10^{-21}$\\
Characteristic age, $\tau$ (yr)\dotfill               & $8\times10^{9}$ \\
Magnetic field, $B_0$ (G)\dotfill                 & $1.9\times10^{8}$ \\
Mass function, $f_1$ (M$_\odot$)\dotfill        & $0.0078962167$\\
Pulsar mass, $m_{1}$ (M$_\odot$)\dotfill        & $1.3(2)$\\
First derivative of $x$, $\dot{x}$ ($10^{-15}$)\dotfill
                                                & 6.7(2)\\
\enddata
\tablenotetext{a}{Figures in parentheses are 68\% confidence uncertainties in the last digit quoted.}
\tablenotetext{b}{Keplerian orbital elements $P_b$, $x$, $e$, $T_0$, and $\omega$ are covariant with 
post-Keplerian elements $\cos i$, $\Omega$, and $m_2$.  This covariance is not reflected in the values
of Keplerian elements and uncertainties quoted in this table, which were derived
in a timing model with $\cos i$, $\Omega$, and $m_2$ fixed to their best-fit values.}
\end{deluxetable*}

\section{Timing Analysis: Parameter Values}\label{sec:grid}

We fit the measured TOAs to a timing model incorporating
the phenomena described in \S\ref{sec:narrative}.  
We used a hybrid procedure to determine the timing parameters
and their uncertainties.  Standard least-squares methods are 
adequate for fitting most of the quantities in the timing model.
However, for the orientation and mass parameters,
$i$, $\Omega$, and $m_2$, 
the $\chi^2$ surfaces are not ellipsoidal in
the parameter space of interest.  To investigate the allowed
ranges of these parameters (and, ultimately, the other parameters
as well), we analyzed timing solutions over a uniformly
sampled three dimensional grid of trial values of
$\cos i$, $m_2$, and $\Omega$, in the vicinity of the
$\chi^2$ minimum.  For each combination of $\cos i$, $m_2$,
and $\Omega$, we calculated  $\dot{x}$, the Shapiro delay parameters, 
and the annual-orbital perturbation corrections, according to
equations~\ref{eqn:shapiro}, \ref{eqn:xdot}, 
\ref{eqn:annorbx}, and \ref{eqn:annorbomega}.  We then performed
a timing fit, in which these quantities held fixed while all other
parameters were allowed to vary.  We recorded the resulting values of $\chi^2$
for each grid point.

The minimum $\chi^2$ timing solution is at $\cos i=0.31$, 
$m_2=0.28$\,M$_\odot$, and $\Omega=87^\circ$.
The parameters corresponding to this solution
are summarized in Table~\ref{tab:param}.  Our results
show good agreement with the less precise
results of \cite{vb03}.

We used a statistically rigorous procedure to calculate
probability distribution functions (PDFs) for the individual
quantities $\cos i$, $m_2$, $\Omega$, and $m_1$ and their
corresponding uncertainties in table~\ref{tab:param}.
The procedure is a straightforward
three-dimensional extension of the Bayesian algorithm described in
appendix A of \cite{sna+02}.   We assigned a probability to
each grid point based on the difference between its $\chi^2$
and the minimum $\chi^2$ on the grid. 
After suitable normalization, we summed the 
probabilities of all points associated with a given range
of $m_1$ (or $m_2$, $\cos i$, or $\Omega$) to calculate 
the PDF.  

In effect,
we are incorporating a uniform prior distribution
in $\cos i$, $\Omega$, and $m_2$.  The uniform distributions
in $\cos i$ and $\Omega$ correspond to the probability distributions
for the angular momentum vectors of randomly oriented orbits, while
the uniform distribution of $m_2$ is simply an ad hoc assumption.

The calculated
PDFs are given in Figure~\ref{fig:pdfs}.  They yield the
68\% confidence estimates given in Table~\ref{tab:param},
most notably 
$m_1=1.3\pm0.2\,\msun$ 
and $m_2=0.28\pm 0.03\,\msun$.  
The 95\% confidence estimates on the masses are
$m_1=1.3_{-0.3}^{+0.4}\,\msun$ and
$m_2=0.28^{+0.06}_{-0.04}\,\msun$.

\begin{figure}[b]
\plotone{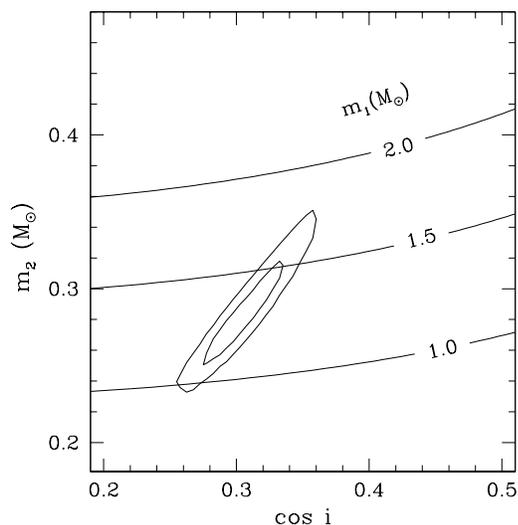}
\caption{Contours of 68\% and 95\% confidence intervals
in $\cos i-m_2$ space.}
\label{fig:cosim2} 
\end{figure}

\begin{figure}[b]
\plotone{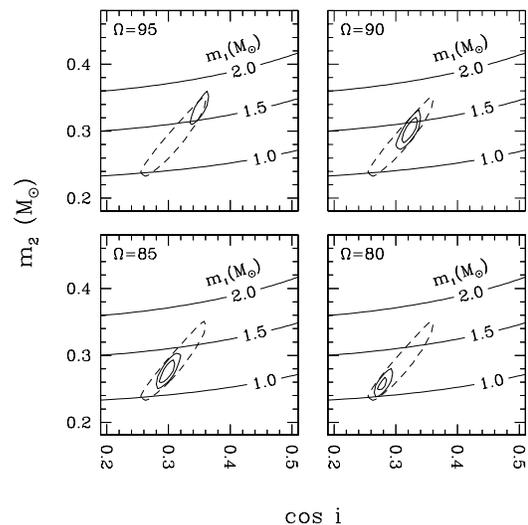}
\caption{Contours of 68\% and 95\% confidence intervals
in $\cos i-m_2-\Omega$ space.  Four slices of constant $\Omega$
are shown; in each, the intersection of the slice with
the 68\% and 95\% confidence region is given.  The
slice at $\Omega=95^\circ$ intersects only the 95\% region.
The dashed line corresponds to the 95\% confidence region
after marginalizing over all values of $\Omega$ (cf.
Figure~\ref{fig:cosim2}).}
\label{fig:slices}
\end{figure}

The interdependence between these parameters is show in the
multidimensional confidence plots in Figures~\ref{fig:cosim2}
and \ref{fig:slices}.  To produce Figure~\ref{fig:cosim2}, we
summed normalized probabilities across all values of $\Omega$
to yield a two dimensional probability distribution in $\cos i$
and $m_2$.  The contours enclose 68\% and 95\% of the probability
distribution in this space.  In Figure~\ref{fig:slices} we
show representative ``slices'' in constant $\Omega$ of the
regions enclosing 68\% and 95\% of the three dimensional grid.

\section{Discussion}

\subsection{Orbital Period--Core Mass Relation}

\subsubsection{Testing the Relation}

\begin{figure}[b]
\plotone{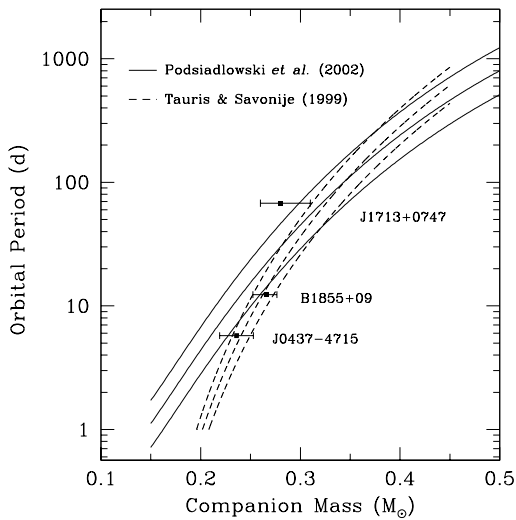}
\caption{Theoretical $m_2-P_b$ relation and measured 
white dwarf masses in pulsar--white dwarf binaries.  
Measured masses are from \cite{vbb+01} (J0437$-$4715),
\cite{ktr94} (B1855+09), and this work (J1713+0747).}
\label{fig:pbm2}
\end{figure}

Binary evolution theory predicts a specific relationship
between the orbital period, $P_b$, of an evolved neutron star--white
dwarf binary and the secondary mass, $m_2$, which is presumed 
to be equal to the helium core of the progenitor of the secondary.
Measured values of $m_2$ can be used to test this relation.
Figure~\ref{fig:pbm2} shows the relations derived from
binary evolution tracks calculated by \cite{ts99a} and
\cite{prp02b}.  The latter curves are the median and upper
values of $m_2$ of \cite{rpj+95}, which provide good bounds
on the tracks of \cite{prp02} (P. Podsiadlowski, private
communication). The models are in good agreement
with the measured values.

\subsubsection{Implications of the Relation for PSR~J1713+0747 System Masses}

If, rather than using timing measurements to test the orbital period--core
mass relation, we accept this relation as correct, than we can use it to 
refine
the range of pulsar mass allowed by the
timing data.  For the orbital period of J1713+0747, \cite{ts99a}
calculate the secondary mass to be $0.31\,{\rm
M}_\odot<m_2<0.34\,\msun$, with the lower end of this range
for a population I progenitor and the higher end for a population II
progenitor.  \cite{prp02} give a similar range, $0.30\,{\rm
M}_\odot<m_2<0.35\,\msun$ (again using the higher end of the
range of \citealt{rpj+95}).  

Repeating the statistical analysis of timing solutions on a
$\cos i-m_2-\Omega$ as in \S~\ref{sec:grid},  but restricting
$m_2$ to the range $0.30\,{\rm M}_\odot<m_2<0.35\,{\rm M}_\odot$
yields a value of $m_1=1.53_{-0.06}^{+0.08}$ (68\% confidence).

\subsection{Neutron Star Mass}

Masses of pulsars and secondary stars in neutron star--neutron
star binaries all fall within the range 1.25 to 1.44\,\msun.
This is near the minimum mass for
neutron star formation; the maximum mass is not
known, and may range up to 3\,\msun\ \citep[see][for a review]{lp04}.
Because binary systems which evolve into pulsar--white dwarf binaries like
the PSR~J1713+0747 system undergo extended
periods of mass transfer, the pulsars in these systems
might be expected to be heavier than those in neutron star--neutron
star binaries.

\Citet{tc99} studied the 
entire binary pulsar population and found it to be
consistent with a narrow Gaussian distribution with mean
1.35\,\msun\ and width 0.04\,\msun.  Their calculation
included a statistical analysis of the pulsar--white
dwarf population, under the assumption that the binaries
were randomly oriented in space and that the orbital period-core 
mass relation held, and they found that these pulsars
could be drawn from the same mass distribution as pulsars
in neutron star--neutron star binaries.

The measured mass of PSR~J1713+0747 based on
observations alone, $m_1=1.3 \pm0.2\,\msun$, 
is in excellent agreement with the 1.35\,\msun\ value.  
In contrast, the mass derived when
the orbital period-core mass constraint is imposed,
$m_1=1.53_{-0.06}^{+0.08}\,\msun$, implies a significantly
heavier neutron star, perhaps having accreted
$\approx 0.2$\,\msun\ if initially formed at $1.35$\,\msun.
There are two other pulsar--helium white
dwarf binaries with well measured Shapiro delays, 
PSRs~J0437$-$4715 and B1855+09, which have pulsar
masses $1.58\pm 0.18\,\msun$ and $1.50^{+0.26}_{-0.14}\,\msun$, respectively
\citep{vbb+01,ktr94}\footnote{A separate analysis of some
of the Kaspi et al. (1994) data by \cite{tc99} reported a smaller
mass, $1.40\pm0.10\,\msun.$}.   On the other hand, 
PSR~J2019+2425, in a similar system, shows a
lack of a detectable Shapiro delay and observed secular changes
which imply an upper limit of 1.51\,\msun\ and a median likelihood
value of 1.33\,\msun\ \citep{nss01}.  The uncertainties on
all of these measurements are frustratingly large;
the question of the distribution of pulsars masses
in these systems remains open.

\subsection{Parallax, Distance, and Velocity}

Because the solar wind introduces annual perturbations on the
pulse arrival times, the solar wind electron density parameter,
$n_0$, is highly covariant with parallax, $\pi$, and position,
$\alpha$ and $\delta$, in the pulse arrival time model.
For $\pi$, the measurement uncertainty using any fixed solar
wind model is $\pm0.05$\,mas, while the uncertainty due to
the poor constraint on $n_0$ is $\pm0.06$\,mas.  We combine
these uncertainties in quadrature to find $\pi=0.89\pm0.08$\,mas.
As a check, a special timing fit was done incorporating
only days on which multifrequency observations were made,
and fitting for a separate value of DM on each day on which
observations were made, yields a parallax of $0.93\pm0.24$,
consistent with our preferred value.  The parallax measurement
corresponds to a distance of $d=1.1\pm 0.1$\,kpc.

For PSR J1713+0747's measured value of DM, the NE2001 model of the Galactic 
electron density \citep{cl02} gives a distance of 0.9\,kpc,
in good agreement with the parallax measurement.

Right ascension and declination are also highly covariant with $n_0$;
their uncertainties quoted in Table~\ref{tab:param} 
were calculated in the same manner as the parallax uncertainty.

It is well established that millisecond pulsars
have substantially smaller velocities than
the bulk pulsar population, with
a mean transverse velocity of 
$85 \pm 13$\,km\,s$^{-1}$
\citep{cc97,tsb+99,nt95}.  
The transverse velocity of PSR~J1713+0747, 
$V = \mu d = 33 \pm 3 \rm{\,km\,s^{-1}}$,
is small even by the
standards of millisecond pulsars.

\subsection{Kinematic Corrections to Spin Period Derivative;
Age \& Magnetic Field}

The observed pulse period derivative, $\dot{P}_{\rm obs}=-\nu_1/\nu_0^2$,
is biased away from its intrinsic value, $\dot{P}_{\rm int}$,
by Doppler accelerations.  \cite{dt91}
analyzed this bias
for orbital period derivatives;  their work applies
equally well to spin period derivatives.
The observed and intrinsic quantities are related
by 
\begin{equation}
\dot{P}_{\rm int} = 
  \dot{P}_{\rm obs} 
 - \Delta\dot{P}_{\rm PM} 
 - \Delta\dot{P}_{\rm rot} 
 - \Delta\dot{P}_{\rm z}, 
\end{equation}
where 
$\Delta\dot{P}_{\rm PM}=P\mu^2 d/c$ is the bias due to proper motion;
$\Delta\dot{P}_{\rm rot}=P({\bf a}_{\rm rot}\cdot{\bf n})/c$ is the
bias due to the relative acceleration of pulsar
and Earth in the Galactic plane, ${\bf a}_{\rm rot}$, projected into a unit vector
pointing from Earth to pulsar, ${\bf n}$; and
$\Delta\dot{P}_{\rm z}=P({\bf a}_{\rm z}\cdot{\bf n})/c$ is the bias due to
the acceleration of the pulsar toward the galactic
disk, ${\bf a}_z$, projected into the line-of-sight.

For PSR~J1713+0747, we find
$\Delta\dot{P}_{\rm PM}=0.46\times 10^{-21}$, 
$\Delta\dot{P}_{\rm rot}=0.22\times 10^{-21}$ and
$\Delta\dot{P}_{\rm z} = -0.32\times 10^{-21}$, the latter
calculated using the potential of 
\cite{kg89}.  The net bias is
$\Delta\dot{P}=0.44\times 10^{-21}$, so
that we estimate the intrinsic period derivative
to be $\dot{P}_{\rm int}=8.1\times 10^{-21}$.

The characteristic age of PSR~J1713+0747
is 
$ \tau = P/2 \dot{P}_{\mathrm{int}} = 
         8 \mathrm{\,Gyr}$.   This
probably overestimates the true age of the pulsar.
Hansen \& Phinney (1998)\nocite{hp98b} 
obtain a considerably lower estimate 
by using the optical measurements of
Lundgren, Foster, \& Camilo (1996)\nocite{lfc96} to
infer the
age of the white dwarf to be 6.3\,Gyr $< \tau <$ 6.8\,Gyr.

A discrepancy between the characteristic
age and the true age of a millisecond pulsar
is not uncommon \citep[e.g.,][]{nt95}, and
implies that the millisecond pulsar period immediately
after spin-up was close to its present day value.  It
seems likely that millisecond pulsars form
with periods of a few milliseconds \citep{bac98}.
 
The surface 
magnetic field strength of the pulsar according
to conventional assumptions is
$ B_0 = 3.2 \times 10^{19} (P \dot{P}_{\mathrm{int}})^{1/2} 
\mathrm{\,G}
= 1.9 \times 10^8 \mathrm{\,G}$.
This value is typical of millisecond pulsars.

\subsection{Timing Noise}

The PSR~J1713+0747 timing data show timing noise.  Analysis
of the timing noise is complicated by the large gap in
the data between 1994 and 1998 
and because of the need to allow an 
offset in the arrival times before and after the gap.  
To analyze timing noise, we calculated residuals for a timing model
fitting for only the pulsar spin-down parameters $\nu_0$
and $\nu_1$ and an arbitrary offset between the pre-1994 and
post-1998 data.  Astrometric and binary parameters were held
fixed at the values derived from the whitened timing model fit.
The results are shown in Figure~\ref{fig:res}a.

Timing noise can be quantified by the fractional stability statistic 
$\sigma_z$, an adaptation of the ``Allan variance'' statistic 
with modifications appropriate for pulsar timing
\citep{mte97}.  In essence, $\sigma_z$ for time interval $\tau$
is calculated by dividing the post-fit residual arrival times, $r(t)$,
into intervals spanning length $\tau$, and fitting third
order polynomials over each interval,
$r(t) = \Sigma_{i=0}^{3} c_i(t\!-\!t_0)^i$.  The first three terms
of the polynomial are accounted for by the pulsar spin-down model.
The final term, $c_3$, is a measure of timing noise.  The statistic
is calculated by appropriately scaling the root-mean-square
value of $c_3$ from all intervals of a given length:
$\sigma_z(\tau)=(1/2\sqrt{5})\tau^2\left<c_3^2\right>^{1/2}$.

\begin{figure}[b]
\plotone{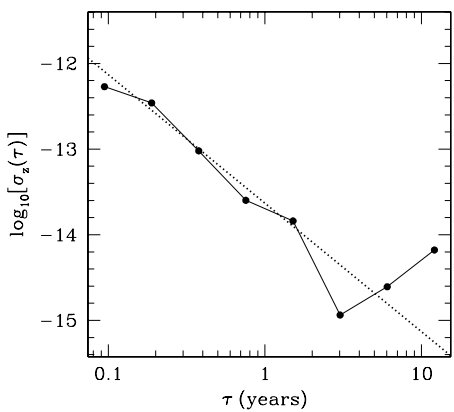}
\caption{Noise statistic $\sigma_z$ as a function
of time scale, $\tau$.  Gaussian noise would result in
$\sigma_z \sim\tau^{-3/2}$, parallel to the dashed line.
The upturn at $\tau>2\,$yr is indicative of timing noise
with a red spectrum.  The $\sigma_z$ values at the longest
time scales may be affected by the presence of a 4-year gap
in the data and by the need for an arbitrary offset between
pre- and post-gap data; however, we believe the upward trend at
$\tau>2$\,yr to be qualitatively correct.
}
\label{fig:sigmaz}
\end{figure}

Figure~\ref{fig:sigmaz} shows $\sigma_z(\tau)$ of PSR~J1713+0747
as calculated from the residual pulse arrival times 
shown in Figure~\ref{fig:res}a. For uniformly
sampled data exhibiting white noise, $\sigma_z$
would fall off as $\tau^{-3/2}$;  this is the behavior
observed on all but the longest time scale.
At $\tau=12$\,yr, timing noise is evident.
Because of the arbitrary offset fit between the 1994
and 1998 data, it is possible that the calculated
value of $\sigma_z(\tau)$ underestimates its true value
on the longest time scale.

The physical mechanism underlying timing
noise is not known.
\citet{antt94} studied 
a large collection of pulsars using a statistic, $\Delta_8$,
which is the logarithm of a scaled version of $\sigma_z(\tau)$ 
at $\tau=10^8\,{\rm s}\approx 3\,{\rm yr}$
\citep[see also][]{bac04}.  From Figure~\ref{fig:sigmaz},
it is evident that at a time scale of 3\,yr, $\sigma_z$ is
dominated by measurement noise, so the measured values
are upper limits on the intrinsic irregularities in the
pulsar signal.  Estimating $\log\sigma_z(3\,{\rm yr})\lesssim -14.2$
for PSR~J1713+0747,
we calculate $\Delta_8\lesssim -5.6$.  \citet{antt94} found $\Delta_8$
to be correlated with pulse period derivative according
to $\Delta_8=6.6+0.6\log \dot{P}$.  
This formula predicts
$\Delta_8=-5.5$ for PSR J1713+0747, close to the observed
upper limit.  Thus the rotational stability of J1713+0747 is as
good (or better) than expected, despite the long-term timing
noise.

\subsection {Solar System Ephemerides}\label{sec:eph}

\begin{figure}[b]
\plotone{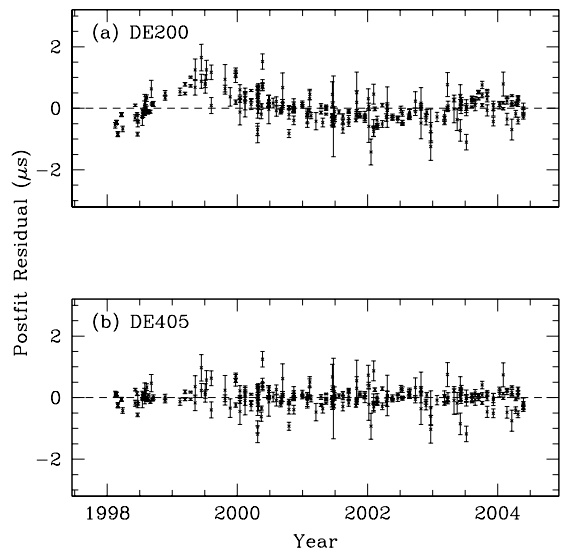}
\caption{Residual arrival times from
timing models for 1998-2004 data 
calculated using (a) the DE200 solar system ephemeris
and (b) the DE405 solar system ephemeris.
\label{fig:eph}}
\end{figure}

As discussed above, we used the DE405 solar system ephemeris
to reduce the pulse arrival times to the solar system 
barycenter.  To reduce pulse arrival times measured 
with uncertainty of only $\sim 200$\,ns, the Earth's
position with respect to the barycenter must be known
with precision $60$\,m.  We analyzed our data with both
DE405 and its predecessor, DE200.  To compare the two
ephemerides, we performed timing fits on the the post-1998
data.  Excluding the earlier data from these tests allowed
us to avoid problems stemming from the arbitrary offset
between the earlier and later data.  Using each ephemeris,
we fit the data to a full timing model, fitting for
all the standard parameters but not allowing pulse frequency
derivatives above $\nu_1$ (i.e., no timing noise terms).  
The results are shown in Figure~\ref{fig:eph}.   
The DE405 ephemeris clearly gives a better fit.
The differences in timing quality
are easily explained by differences
in the ephemerides themselves, particularly the incorporation 
of improved measurements of outer planet masses into DE405
(E. M. Standish, private communication).

\subsection {Testing Strong Field Gravity}\label{sec:sep}

PSR~J1713+0747 is one of several long-orbit, low-eccentricity binary
pulsars which have been used to set limits on violations
of equivalence principles \citep{ds91,wex97,bd96,wex00}.  The relevant
observational signatures depend on the orientations of the binary
systems under study, something not usually known; hence, probabilistic
arguments have been used to constrain equivalence principle violations
based on observations of several pulsars.  In contrast, we have
established
the orientation ($i$ and $\Omega$) to PSR~J1713+0747, and we have
measured
its distance as well.  This allows us to set the first absolute
limits both on violation of the Strong Equivalence Principle (SEP) and
on the magnitude of the strong-field equivalent of the Parametrized
Post-Newtonian (PPN) parameter $\hat \alpha_3$.

\subsubsection {The Strong Equivalence Principle}

If the SEP were violated, objects with different fractional mass
contributions from self-gravitation would fall differently in an
external gravitational field.  This is quantified by the
parameter $\eta$, defined through $m_g/m_i=1+\eta E_g/mc^2$,
where $m_g$, $m_i$, and $E_g$ are the gravitational mass, 
inertial mass, and gravitational self-energy of a body.
Nonzero values of $\eta$ would result in the polarization of binary
orbits \citep{nor68b}.  Lunar laser ranging experiments set limits on
the polarization of the Earth-Moon orbit in the gravitational
field of the Sun, constraining $|\eta|$ to be less than 0.001
\citep{dbf+94,wil01}.  In the case of a pulsar--white-dwarf binary,
the orbit would be polarized in the direction of the gravitational
pull of the Galaxy.  The parameter to be constrained, 
$\Delta$, is similar to $\eta$, but without the
requirement of linear dependence on $E_g/mc^2$.
It is defined for
an individual body, $a$, by $(m_g/m_i)_a = 1+\Delta_a$;
dynamics of a binary orbit depend on the difference
$\Delta=\Delta_1-\Delta_2$
between the two objects \citep[see][for full details]{ds91}.

The (small) ``forced'' eccentricity ${\bf e}_F$ of the orbit induced
by the SEP violation may be written as:
\begin{equation}
|{\bf e}_F| = \frac{1}{2}\frac{\Delta {\bf
g}_{\perp}c^2}{FG\,(m_1+m_2) (2\pi/P_{\rm b})^2},
\label{equation:eforced}
\end{equation}
where ${\bf g}_{\perp}$ is the projection of the Galactic
gravitational field
onto the orbital plane and, in General
Relativity, $F = 1$.  This forced
eccentricity may be comparable in magnitude to the ``natural''
eccentricity ${\bf e}_R(t)$ of the system, which by definition rotates
at the rate of the advance of periastron and may, at any time, be
oriented in such a way as to nearly cancel the forced eccentricity.
Defining the angle between these two vectors as $\theta$,
\citet{wex97} writes the inequality:
\begin{equation}
|{\bf e}_F| \leq e \xi_1(\theta), \,\,\,\,\,
\xi_1(\theta) = \left\{ \begin{array}{r@{\quad:\quad}l}
        1/\sin\theta & \theta \in [0,\pi/2) \\
        1 & \theta \in [\pi/2,3\pi/2] \\
        -1/\sin\theta & \theta \in (3\pi/2,2\pi)
        \end{array}  \right. ,
\label{equation:xi1}
\end{equation}
where $e$ is the observed eccentricity.  

The projection of ${\bf g}$ onto the orbital plane can be written as
\citep{ds91}:
\begin{equation}
|{\bf g}_{\perp}| = |{\bf g}|[1-(\cos i \cos\lambda + \sin i
\sin\lambda\sin(\phi_{\rm g} - \Omega))^2]^{1/2},
\label{equation:gperp}
\end{equation}
where $\phi_{\rm g}$ is the angle of the projection of ${\bf g}$ onto
the plane of the sky (effectively a celestial position angle, defined
north through east as for $\Omega$), and $\lambda$ is the angle
between ${\bf g}$ and the line from the pulsar to the Earth.  The value of
$|{\bf g}|$ can be determined from models of the Galactic potential
\citep[e.g.,][]{kg89} and the Galactic rotation curve; for a pulsar
out of the Plane of the Galaxy, $\lambda$ is given by
\begin{displaymath}
\cos \lambda = \frac{-R_0^2+d^2+R_1^2+z^2}{2\,d\,\sqrt{R_1^2+z^2}},
\label{equation:coslambda}
\end{displaymath}
where $R_1 = (R_0^2+(d \cos b)^2 -2 R_0 d \cos b \cos l)^{1/2}$ is
the Galactic radius of the pulsar, $R_0$ is the distance from the
Earth to the Galactic center, $z$ is the distance from the pulsar
to the Galactic plane, and $l$ and $b$ are the Galactic coordinates
of the pulsar.

Historically, both $\theta$ and $\Omega$ have been completely unknown
for the pulsars used to test for violation of the SEP, and the tests
have therefore made statistical arguments, assuming both angles to be
uniformly distributed between 0 and $2\pi$ \citep[e.g.,][]{ds91}.  The
masses have also been poorly constrained and thus averages over likely
model populations were also needed \citep{wex00}.  Following these
procedures, an ensemble of long-orbit pulsars yields a limit of
$|\Delta| < 0.009$ at 95\% confidence.

With both masses and the angle of the line of nodes well-constrained
for PSR~J1713+0747, a single, more robust limit on violation of the
SEP becomes possible.  Because of the complicated dependence of
$\Delta$ on the pulsar distance, as well as the asymmetric
distribution of allowed masses, we obtain the limit via Monte Carlo
simulation, assuming the parallax to be normally distributed and
$\theta$ to be uniformly distributed, and sampling the allowed
$m_2$--$\cos i$--$\Omega$ range according the probability distribution
derived from the $\chi^2$ grid discussed above.  We find that
$|\Delta| < 0.013$ at 95\% confidence.  This is nearly as good a limit
as could be expected from this pulsar, as just over 90\% of the
Galactic acceleration vector is parallel to the plane of the orbit.
The most stringent test of SEP violation, however, continues to rely
on an ensemble of pulsars.

\subsubsection {Post-Newtonian Parameter $\hat \alpha_3$}

The parameter $\alpha_3$ is one of ten PPN parameters
formulated to describe departures from General Relativity in the
weak-field limit \citep{wn72}.  Its strong-field analog, $\hat
{\alpha_3}$, can be tested by pulsar timing
\citep{de92,de96a}.  A non-zero $\hat\alpha_3$ would imply both
violation of
local Lorentz invariance and non-conservation of momentum
\citep[e.g.,][]{wil01}.  For a pulsar--white-dwarf binary, the net 
effect would be to cause an acceleration of the binary system given by
\citep{bd96}:
\begin{equation}
{\bf a}_{\hat \alpha_3} = \frac{1}{6}\hat \alpha_3 c_{\rm p}{\bf w}
\times {\bf \Omega}_{\rm Sp},
\label{alpha3accel2}
\end{equation}
where ${\bf \Omega}_{\rm Sp}$ is the spin angular frequency of the
pulsar ($|\Omega_{\rm Sp}|=2\pi\nu_0$), ${\bf w}$ is the absolute velocity of
the system, and $c_{\rm
p}$ denotes the compactness of the pulsar, roughly the fraction of its
mass contained in gravitational self-energy.  An approximate
expression for the compactness is $c_{\rm p} \simeq
0.21\,m_1/M_{\odot}$ \citep{de92}.  As above, this acceleration will
induce a ``forced'' eccentricity in the orbit, given by
\begin{equation}
|{\bf e}_F| = \hat \alpha_3\frac{c_{\rm p}|{\bf w}|}{24\pi}\frac{P_{\rm
 b}^2}{P}\frac{c^2}{G(m_1+m_2)}\sin\beta,
\label{equation:efalpha3}
\end{equation}
where $\beta$ is the angle between ${\bf w}$ and ${\bf \Omega}_{\rm
Sp}$, and $P$ is the spin period of the pulsar: $P=2\pi/\Omega_{\rm
Sp}$.  Using an ensemble of pulsars and assuming random distributions
of the orbital inclinations, \citet{wex00} found a limit of $|\hat
\alpha_3| < 1.5 \times 10^{-19}$.  To calculate the limit imposed by
PSR~J1713+0747, we again proceed via Monte Carlo simulation, sampling
the parameters as described above.  We assume that the pulsar spin and
orbital angular momenta have been aligned during the spin-up episode
that recycled the pulsar.  Thus with the full orientation of the orbit
known for each simulation point, we can easily calculate $\beta$ given
any ${\bf w}$.  This absolute velocity is taken to be in the reference
frame of the Cosmic Microwave Background (CMB), and is calculated from
the motion of the Solar System in the CMB reference frame (369 km/s
toward $\alpha=11\fh 3$, $\delta=-7^{\circ}$;
\citealt{fcc+94}) and the three dimensional velocity of the binary 
system relative to the Solar System.  The radial component of this
second velocity vector
is unknown.  We checked a range of radial velocities
from $-200$\,km/s and $+200$\,km/s; for each simulation
point, we adopted the value that gives the smallest projection of the
total
velocity onto the plane of the orbit. This value was always in the
range $-60$ to +25\,km/s.  We thus arrived at an absolute 95\%-confidence
limit of $|\hat \alpha_3| < 1.2 \times 10^{-19}$.  This is
better than the limit derived from the
ensemble of pulsars and, as it is less statistical in nature, it may
be considered a
robust true limit on $\hat{\alpha_3}$.  Future refinement of the timing
parameters will improve this limit somewhat, with the floor ultimately
determined by the orbital geometry.

\section{Conclusion}

PSR~J1713+0747 has lived up to its promise as one of the best
pulsars for high precision timing, with 200\,ns residual pulse
arrival times attained on time scales of years, and residuals
well under 2$\mu$s over the full 12-yr data set.
Major findings include:

1.  The pulsar mass is constrained to $m_1=1.3 \pm0.2\,\msun$
by the measured Shapiro delay.  If the secondary mass is
restricted to values predicted by the theoretical
orbital period-core mass relation, the pulsar mass
is somewhat higher, $m_1=1.53_{-0.06}^{+0.08}$.

2.  The parallax is $\pi=0.89\pm0.08$\,mas, corresponding to
a distance of $1.1\pm0.1$\,kpc.  This is consistent with
predictions based on the pulsar's dispersion measure.

3.  The orientation of the binary has been fully determined
by the combined measurements of Shapiro delay and of perturbations
of orbital elements due to relative Earth-pulsar motion.
The orientation and very low ellipticity of the orbit
lead to an improved constraint
on deviations from general relativity.

The timing precision attainable on short time scales
appears to be limited by measurement precision.  There
is every reason to expect timing precision to improve in the coming
years as a new generation of wide bandwidth coherent
dedispersion systems are employed at radio telescopes,
directly resulting in more precise measurement of the
Shapiro delay and hence of the pulsar and white
dwarf masses.

\acknowledgements

The Arecibo Observatory is a facility of the National Astronomy
and Ionosphere Center, operated by Cornell University under a
cooperative agreement with the National Science Foundation.
DJN is supported by NSF grant AST-0206205. DCB acknowledges NSF
grants AST-9731106 for development of ABPP and AST-9820662
and AST-0206044 for support of his scientific program.
IHS holds an NSERC UFA and is supported by a Discovery Grant.
We thank F. Camilo for supplying the data presented in
\cite{cfw94} and for comments on the manuscript; 
K. Xilouris for coordinating
early post-Arecibo-upgrade observations; R. Ferdman for
help collecting data; B. Reid for analysis of Mark\,IV pulse
profiles; and R. Kipphorn for analysis of some ABPP data.


\end{document}